\documentclass[%
 reprint,
superscriptaddress,
 amsmath,amssymb,
 aps,
]{revtex4-2}

\usepackage{dsfont}
\usepackage[export]{adjustbox}
\usepackage{ragged2e}
\usepackage{amsmath}
\usepackage{amsfonts}
\usepackage{amssymb}
\usepackage{xcolor}
\usepackage{graphicx}
\usepackage{subfiles}
\usepackage[colorlinks,linkcolor=red,citecolor=blue]{hyperref}
\usepackage{subcaption}
\usepackage{graphicx}
\usepackage{dcolumn}
\usepackage{bm}
\usepackage{amssymb,amsmath,color}
\usepackage{booktabs}
\usepackage{amsthm}
\usepackage{dsfont}
\usepackage{tcolorbox}
\usepackage{cases}
\usepackage{physics}
\usepackage{subfiles}
\newcommand\numberthis{\addtocounter{equation}{1}\tag{\theequation}}

\newenvironment{thisnote}{\par\color{black}}{\par}

\theoremstyle{definition}

\newcommand{\bc}{\begin{center}}
\newcommand{\ec}{\end{center}}

\def\blk{\color{black}}

\begin{document}


\def\br{\biggr}
\def\bl{\biggl}
\def\Br{\Biggr}
\def\Bl{\Biggl}
\def\be\begin{equation}
\def\ee{\end{equation}}
\def\bea{\begin{eqnarray}}
\def\eea{\end{eqnarray}}
\def\f{\frac}
\def\n{\nonumber}
\def\l{\label}

\title{Optimal Quantum Control of Charging Quantum Batteries}

\author{R. R. Rodr\'iguez}
\email{ricard.ravellrodriguez@phdstud.ug.edu.pl}
\address{International Centre for Theory of Quantum Technologies, University of Gdansk, Jana Bażyńskiego 1A, 80-309 Gdansk, Poland}
\author{B. Ahmadi}
\email{borhan.ahmadi@ug.edu.pl}
\address{International Centre for Theory of Quantum Technologies, University of Gdansk, Jana Bażyńskiego 1A, 80-309 Gdansk, Poland}
\author{G. Su\'arez}
\address{International Centre for Theory of Quantum Technologies, University of Gdansk, Jana Bażyńskiego 1A, 80-309 Gdansk, Poland}
\author{P. Mazurek}
\address{International Centre for Theory of Quantum Technologies, University of Gdansk, Jana Bażyńskiego 1A, 80-309 Gdansk, Poland}
\author{S. Barzanjeh}
\address{Department of Physics and Astronomy, University of Calgary, Calgary, AB T2N 1N4 Canada}
\author{P. Horodecki}
\address{International Centre for Theory of Quantum Technologies, University of Gdansk, Jana Bażyńskiego 1A, 80-309 Gdansk, Poland}

\begin{abstract}
Quantum control allows us to address the problem of engineering quantum dynamics for special purposes. While recently the field of quantum batteries has attracted much attention, optimization of their charging has not benefited from the quantum control methods. Here we fill this gap by using an optimization method. We apply for the first time this convergent iterative method for the control of the population of a bipartite quantum system in two cases, starting with a qubit-qubit case. The quantum charger-battery system is considered here, where the energy is pumped into the charger by an external classical electromagnetic field. Secondly, we systematically develop the original formulation of the method for two harmonic oscillators in the Gaussian regime. In both cases, the charger is considered to be an open dissipative system. Our optimization takes into account experimentally viable problem of turning-on and off of the charging external field. Optimising the shape of the pulse significantly boosts both the power and efficiency of the charging process in comparison to the sinusoidal drive. The harmonic oscillator setting of quantum batteries is of a particular interest, as the optimal driving pulse remains so independently of the temperature of environment.
\end{abstract}

\maketitle
\section{Introduction}

Energy processing in quantum systems attracts attention for both fundamental and practical reasons. While second law of thermodynamics holds in its generalised form \cite{Strasberg,brandao2015second,Esposito_2011,Lobejko_2021}, relinquishing classical constrains in microscopic thermal machines may lead to boost in their power \cite{Uzdin1,Uzdin2,Klatzow}. Potential for quantum advantage was one of the factors (apart from sheer curiosity) motivating research in microscopic thermal machines, both from the quantum open system \cite{Kosloff,Gelbwaser,Bohr_Brask_2015,Correa_2014,Dann_2020,ghosh2021fast} (see Ref. \cite{Bhattacharjee2021} for a comprehensive review) and resource theory \cite{Ng_2018,Sparaciari_2017,Shi_2020,Lobejko_2020,biswas2022extraction} perspectives. Proposals of their realisations \cite{Campisi_2015,Zhang,Gelbwaser2014} were recently followed by experimental implementations in trapped ion systems \cite{Kilian_2016,Maslennikov_2019,Lindenfels}.

Quantum battery \cite{Bhattacharjee2021} is a system which stores useful energy (work), and as such may be integrated into a thermal machine, which operates with the aim of charging the battery. Alternatively, quantum batteries are of independent interest as reservoirs of energy which can be stored and released on demand. Possible application of these systems is to provide energy for operations on low temperature quantum systems from within, as an alternative to energy transfer from exterior sources, leading to noise. Quantum effects in operating quantum batteries have been identified in many body systems, where collective effects enable advantage in charging power \cite{Alicki2013,Hovhannisyan2013,Binder_2015,Campaioli2017}, similarly to metrological settings. Spin chains \cite{Le}, superconducting qubits and quantum dots \cite{Ferraro_2018}, disordered chains \cite{Caravelli,Zhao}, and qubits in an optical cavity \cite{Andolina2019} were all investigated from the perspective of their use as quantum batteries. 

In this paper, we tackle a problem of identifying the optimal classical drive which charges quantum battery in a typical charging setting. Namely, we intend to maximize useful energy stored in the battery, while at the same time reducing energy spent during the charging. We will verify the performance of the method both for qubits and harmonic oscillators, as both are common models of quantum batteries \cite{Andolina2018,Andolina2019}. The proposed method for drive optimization will take into account noise present in the realistic charging scenarios, described within the open system approach.

\begin{figure}
    \includegraphics[width = 0.4 \textwidth]{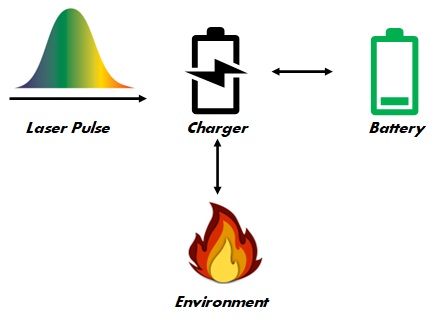}
    \caption{Battery charging: a quantum system, battery, is coupled to another quantum system, charger. The charger is pumped by classical electromagnetic field while interacting with environment. Goal: taking into account all the couplings, optimize over pulse shape such that the battery achieves high energy and ergotropy at a given time. 
    }
    \label{fig:general}
\end{figure}

We will address a charging scenario in which the battery is coupled to an additional quantum system, called charger (see Fig. \eqref{fig:general}). It is the charger which interacts with external laser field, as well as with environment. On one hand, this arrangement  allows for the flow of energy from the field to the battery, while on the other partially isolates it from the effects of noise, improving quality of stored energy. While quantifiers of this quality may vary for particular physical applications, here we are going to use ergotropy \cite{Allahverdyan_2004}. The general setup for battery charging is described in more detail in Sec. \ref{setting}.

Different methods may be used to improve power and quality of charging. An example of a passive method is to exploit qutrit batteries with energy landscape favouring charging through the so-called dark states \cite{Santos2019,Munro2020}, while measurements \cite{Gherardini2020}, empowered by linear feedback protocols \cite{Mitchison_2021,PhysRevApplied.20.044073} have been shown to exhibit improvement with respect to charging which do not take noise into account. Recently, considering noisless battery charging, \cite{Mazzoncini_2023} proposed a method based on Pontryagin's minimum principle to design optimal modulations of the Hamiltonians.    

Here, we develop a passive optimized method for charging quantum batteries of an arbitrary dimension through chargers coupled to classical electromagnetic field and memory-less environment. 
The method relies on an iterative numerical technique \cite{Zhu,Werschnik,Goerz_2019} for optimizing a functional, which in our case encodes the goal state of the battery and the dynamics of the open system, including  noise and external driving. We choose the goal state of the battery to be maximally excited (in case of a qubit) or a specific pure state (in case of a harmonic oscillator battery). The result is a time profile of the electromagnetic field which drives the battery state close to the goal state in a desired time. By a design of the functional, we enforce shape pulses of the field to increase smoothly from zero values at the beginning of the protocol (and vanish in the same manner at the end), which addresses the problem of switching on and off of the laser field, and ensures practicality of obtained solutions. Our method works not only for pure, but also mixed states, which enables us to analyze charging for non-zero temperatures of the environment. The method is presented in Sec. \ref{method}.

When applied to qubit atomic batteries inside an optical cavity (Sec. \ref{qq}), the method leads to battery states of higher energy and ergotropy than the ones previously reported in Ref. \cite{farina2019charger} for experimentally viable parameters. Remarkably, while our method is not fully optimized for minimizing the energy cost of charging, it leads to protocols with lesser energy consumption than the ones reported for qubit batteries. Moreover, in Sec. \ref{oo} we show that the same applies to charging of harmonic oscillator batteries with linear coupling, while in this case optimization over temperature is not needed, provided charger and battery are prepared in the vacuum state.

\section{General setup}\label{setting}

\begin{thisnote}    
We start by arguing for the relevance of the applied model of quantum control. 
\subsection{Battery charging recast as a quantum control problem}
 A desired transformation of a quantum system can be performed in two paradigms: non-autonomous and autonomous. The latter case can be achieved either in continuous dynamical models, as used in the description of thermal machines, or through the application of quantum clocks \cite{Woods2019, Mitchison_2019}. In the latter case, additional quantum systems are prepared and coupled to the main system through time-independent coupling, with the aim of performing a desired operation on the system. This operation may result in activating interactions between parts of the system, and associated energy costs can be calculated by investigating preparation and dynamics of the clock. 
 
 Contrary to this, in the non-autonomous approach, the control on the system is active, with extra care needed when accounting for the energy costs of the desired transformations. This applies both for driving the system through some time-dependent parameter, or in feedback schemes, as they rely on measurements and/or operations performed 'on demand' (feedback schemes can be based both on processing of classical and quantum information, in the spirit of \cite{Lloyd_2000}). 

 Thinking about battery charging as a process in which a system (composed of the battery and the charger) is driven to a desired state, we choose the simplest non-autonomous model of external, classical (laser) drive, with no feedback mechanism present in the protocol. Avoiding autonomous settings is motivated by the simplicity of experimental implementation of a laser drive, while its classical nature is justified for the weak back-reaction imposed on the field by the interaction with the system. Finally, keeping in mind that the goal is to charge the battery with a form of ordered energy (quantified below through its ergotropy), we do not expect the interaction with the environment to leave room for any improvements, which could be achieved through feedback mechanisms (at least for the case of dissipative coupling to thermal baths). Therefore, the classical drive is treated in the open-loop approach, and no information about the state of the system is gathered during the protocol.       

 As mentioned above, to describe the preferred experimental setting, the laser is not to directly interact with the system, and instead a charger is mediating the interaction between the drive and a system on which the transformation needs to be performed. Again, this setting is minimal, and does not allow for investigation of effects which arose though synchronized charging through many chargers. Similar scenarios were recently studied in \cite{Arjmandi_2022}, where, in the absence of classical drive, coherence shared between the oscillator chargers increased the final energy and ergotropy of the battery.

 On the other hand, investigation of systems of multiple batteries driven through a single charger can be in general treated as a direct extension of the single charger/single battery model. 
 
 As we will see in Sec. \ref{oo}, the method remains efficient in designing charging pulses for the setting in which both charger and battery are modelled by harmonic oscillators. This is because, for the assumed linear coupling between the charger and the battery, the dimensionality of the optimization problem can be reduced, as the evolution preserves Gaussianity of the states, which results in good performance of the method. On the other hand, while the method can be applied to a multi-battery case, it requires more computational resources to find the solution which renders advantage both in terms of ergotropy of the battery and energy spent.

  \end{thisnote}

 \subsection{The model}
We move to the detailed description of the single charger/single battery model, driven by a classical laser pulse. We investigate two cases: when both of these systems can be modelled simultaneously as qubits or harmonic oscillators. 

The charger is assumed to dissipate energy into the environment during the charging process. 
The local GKSL master equation \cite{chruscinski2017brief} will be used to describe the dissipative evolution of the whole system. The process starts from the charger in the ground state driven by a classical laser field. The goal is to design an optimized laser pulse such that the final state is achieved at a given time. With the aim to demonstrate efficient charging of qubit and oscillator quantum batteries, we select their final states to be excited and some coherent states, respectively.

Quality of the final state of the battery
is going to be assessed by its ergotropy. Ergotropy is defined as the maximum amount of work that can be extracted from a state $\rho$ in a cyclic unitary process such that the initial Hamiltonian of the system $H(t_0)$ is equal to the final Hamiltonian of the system $H(t_f)$ \cite{Allahverdyan_2004}. Defining $E(\rho):=\tr{\rho H}$ as the internal energy of the system, ergotropy of a state $\rho$ is calculated as
\begin{equation}
\mathcal{E}(\rho):=E(\rho)-E(\rho_p),
\end{equation}
where $\rho_p$ is the passive state associated with $\rho$ by $\rho_p=\min_{U}U\rho U^{\dagger}$, with the minimization over all unitaries. Since the ground state is a passive state only for pure states, in the optimization process we choose the target state of the battery to be an excited pure state to maximize the ergotropy. On the other hand, the energy cost of the drive is calculated as
\begin{equation}\label{Energy-used2}
\mathcal{W}_{\tau}=\int_{0}^{\tau} dt\ |\epsilon(t)|^2,
\end{equation}
where $\tau$ is the final time of the evolution, end $\epsilon(t)$ is the time-dependent amplitude of the drive.

Consequently, to describe efficiency of the control procedure, we will compare the values of extractable ergotropies $\mathcal{E}_{osc}$, $\mathcal{E}_{opt}$  and energy spent $\mathcal{W}_{osc}$, $\mathcal{W}_{opt}$ for the simple, oscillatory drive, and the solution found by the optimization method, respectively. As figures of merit we introduce quality factors 

\begin{equation}
    \alpha_{\mathcal{E}}=: \left(\frac{\mathcal{E}_{opt}(\rho)}{\mathcal{E}_{osc}(\rho)}-1\right) \times 100\%,
\end{equation}

\begin{equation}\label{cost}
    \alpha_{\mathcal{W}}=: \left(\frac{\mathcal{W}_{osc}(\rho)}{\mathcal{W}_{opt}(\rho)}-1\right) \times 100\%
\end{equation}
  for the battery at the final time $\tau$, quantifying how much the optimization improves the charging process in terms of ergotropy extracted and energy spent, respectively, evaluating method's efficiency and efficacy. 

In Refs. \cite{ferraro2018high,farina2019charger,andolina2019quantum,Andolina2019,julia2020bounds} quantum correlations and quantum coherence contained in the initial state of the charger were investigated for possible benefits in transferring energy from quantum chargers to quantum batteries. As specific preparations of the initial state of the charger may cost significant input energy, to have a proper energy accounting we restrict ourselves to product ground states. Note however that the optimized charging method can be applied for an arbitrary initial state of a battery-charger system.

\section{Convergent iterative method for control of the population of a quantum system}\label{method}

The charging process can be described as a transition from the initial state $\hat{\rho}_{0}$ of the system (discharged battery and charger) to a target state $\hat{\rho}^{\text{tgt}}$ (representing battery charged, and an arbitrary state of the charger). The transition results from a dynamical process taking time $\tau$ (we take the initial time to be 0), which, for noisy dynamics, is modelled by the equation of motion:
\begin{equation}
    \label{open-evolution}
    \frac{d\hat{\rho}}{dt}=\mathcal{L}\hat{\rho}
,\hspace{0.3 cm}\hat{\rho}(0)=\hat{\rho}^{0}, \end{equation}
with the initial state $\hat{\rho}^0$ setting a boundary condition. The Liouvillian operator $\mathcal{L}$ takes into account local Hamiltonians, dissipation terms and the classical drive. Under this evolution constraint, driving the system to the desired final state while limiting the energy spent on the drive can be understood through the problem of minimization of the general functional 
\begin{widetext}
\begin{equation}\label{functional}
J\left[\hat{\rho}^{(i)}(\tau),\epsilon^{(i)}(t)\right]=J_{\tau}\left(\hat{\rho}^{(i)}(\tau)\right)+ \int_{0}^{\tau} g_{a}\left(\epsilon^{(i)}(t)\right) \mathrm{d} t,
\end{equation}
\end{widetext}
where $\hat{\rho}^{(i)}(\tau)$ denotes the state at the final time $\tau$ evolved by the field of iteration $i$, $\epsilon^{(i)}(t)$. $J_{\tau}$ is the main part of the functional $J$ and throughout the paper it corresponds to $J_{\tau}=1-\mathcal{F}^{i}(\tau)$, where
$\mathcal{F}^i(\tau)=
\operatorname{tr}\left[\hat{\rho}^{\operatorname{tgt} \dagger} \hat{\rho}^i(\tau)\right]$ is the fidelity between the pure states $\hat{\rho}^i(\tau)$ and $\hat{\rho}^{\operatorname{tgt}}$. 

The remaining part of the functional attests for the energy cost of the drive: 
\begin{align*}\label{field-running costs}
g_{a}\left(\epsilon^{(i)}(t)\right) &=\frac{\lambda}{S(t)}\left(\epsilon^{(i)}(t)-\epsilon^{(i-1)}(t)\right)^{2}  \\\nonumber
&=\frac{\lambda}{S(t)}\left(\Delta \epsilon^{(i)}(t)\right)^{2},\numberthis
\end{align*}
where $\lambda>0$ is a numerical parameter for the optimization and $S(t) \in [0,1]$ allows for the control of the shape of the pulse. Throughout this work $S(t)$ has the form 
\begin{equation}\label{shape-update}
S(t)=\begin{cases}
\sin^2{\left(\frac{\pi}{2} \frac{t}{t_{\textnormal{on}}}\right)} & \text {for} \quad 0\leq t \leq t_{\text {on }} \\
1 & \text { for } \quad t_{\text {on }} < t < \tau-t_{\text {off }} \\
\sin^2 {\left(\frac{\pi}{2}\frac{(t-\tau)}{t_{\textnormal{off}}}\right)}& \text { for } \quad \tau-t_{\text {off }} \leq t \leq \tau,
\end{cases}
\end{equation}
where $t_{\text {on }}=t_{\text {off }}=0.005\tau$. We take the initial guess for the field in the form 
\begin{eqnarray}\label{S}
    \epsilon^{(i=0)}(t)=S(t)\kappa.
\end{eqnarray}
For a specific case, we choose $\kappa$ large enough to achieve fast convergence. The term (\ref{field-running costs}) quantifies the divergence of the pulse shape at iteration $i$ from the shape in the previous iteration. We select pulse shapes with small energies in the initial iterations, in order to obtain a final pulse shape solutions which is not energetically costly. 

The reasoning for the choices of $S(t)$ and $\epsilon^{(i=0)}(t)$ is the following:
we start with the field guess $S(t)\kappa$ which has small values at initial and final stages of the evolution. Minimization of the functional with $S(t)$ going to 0 in these limits suppresses substantial modification of values of the field in the initial and final time regimes, compared to modification of the values of the field in intermediate times. The result is that the optimized pulse smoothly achieves small values at initial and final times. 
Essentially, the optimization strategy is finding a way to ensure that at each iteration the value of the functional diminishes, i.e.
\begin{equation}\label{convergence-equation}
    J\left[\hat{\rho}^{(i+1)}(\tau),\epsilon^{(i+1)}(t)\right] \leq J\left[\hat{\rho}^{(i)}(\tau),\epsilon^{(i)}(t)\right]. 
\end{equation}
Following the original proof of convergence \cite{Zhu} for closed-system dynamics we see that Eq. $\eqref{convergence-equation}$ is satisfied when the update of the field is 
\begin{equation}\label{field-update}
\Delta \epsilon^{(i)}(t)=
\frac{S(t)}{\lambda} \operatorname{Im}\left[\tr{\hat{\sigma}^{(i-1)}(t)\left(\frac{i \partial \mathcal{L}}{\partial \epsilon(t)}\Bigr|_{(i)}\right) \hat{\rho}^{(i)}(t)}\right].
\end{equation}
where $\hat{\rho}$ and $\hat{\sigma}$ represent the density matrix of the forward and backward states respectively. Above, we used a shorthand for 
\begin{equation}
\frac{\partial \mathcal{L}}{\partial \epsilon(t)}\Bigr|_{(i)}=\frac{ \partial \mathcal{L}}{\partial \epsilon(t)}\Bigr|_{\epsilon(t)=\epsilon^{(i)}(t)}.    
\end{equation}
While in general this derivative has to be numerically calculated, this is not the case for charging protocols investigated in this work, where the corresponding Liouvillians are linear in the control field.  $\hat{\sigma}^{(i-1)}$ are the so-called backward states and they evolve according to \footnote{In general though, $\hat{\sigma}^{(i-1)}_k(\tau)=-\frac{\partial J_{\tau}}{\partial \hat{\rho}_k(\tau)}\Bigr|_{(i-1)}$. For the cases computed in the paper, equation \ref{backward-states} is enough.}
\begin{equation}\label{backward-states}
    \frac{d\hat{\sigma}}{dt}=-\mathcal{L}^{\dagger} \hat{\sigma}, \hspace{0.2 cm} \hat{\sigma}(\tau)=\hat{\rho}^{\operatorname{tgt}},
\end{equation}
where the time goes backwards. Having said that, we can proceed to sketch the algorithm. To simplify the explanation, we will consider that $\frac{\partial \mathcal{L}}{\partial \epsilon (t)}$ does not depend on time. We consider a time grid made of $N+1$ time points with the step between them equal to $dt$. The grid starts at $t=0$ and ends at $t=\tau$ and the states are only defined at those points. We use the notation $\epsilon^{(i=0)}(t)$ for the initial guess of the field. First of all, for the iteration $i$ we generate $\hat{\sigma}^{(i-1)}(t)$ by propagating the backward states $\hat{\sigma}$ using Eq. \eqref{backward-states} with field $\epsilon^{(i)}(t)$ from $t=\tau$ to $t=0$ along the $N+1$ time points. Then, we calculate the update of the field with our initial state $\hat{\rho}_0$ using the discretized version of Eq. \eqref{field-update} as
\begin{equation}\label{discrete-field-update}
\Delta \epsilon^{(i)}(\bar{t})=\frac{S(\bar{t})}{\lambda} \operatorname{Im}\left[\tr{\hat{\sigma}^{(i-1)}(t)\left(\frac{i \partial \mathcal{L}}{\partial \epsilon(t)}\Bigr|_{(i)}\right) \hat{\rho}^{(i)}(t)}\right],
\end{equation}
where instead of computing the update for time $t$ we will do so for time $\bar{t}=t+dt/2$. Doing so, we can solve the apparent contradiction of Eq. \eqref{field-update} where the update of the field is calculated at time $t$ whereas the states used for the computation are also at time $t$ and to get those, we will need to propagate them under the field we are calculating now. Therefore, calculating it at $\bar{t}$, the update only depends on the past information. Actually, Eq. \eqref{field-update} is the continuous limit of Eq. \eqref{discrete-field-update} when $dt\rightarrow 0$.

With this update, we can propagate our states $\hat{\rho}$ under the new field while at the same time we keep updating the field sequentially for all time grid, obtaining $\hat{\rho}^{(i)}(t)$ and $\epsilon^{(i)}(t)$.  In doing this, we extend the field values calculated at points $t+dt/2$ to points $t+dt$, tacitly assuming that discretization of the evolution equations does not lead to rapidly oscillating field (this can always be obtained by decreasing $dt$). After that, we move to the next step, $i+1$, where the guess field will correspond to the updated field of the iteration from before. This procedure keeps repeating until the convergence of the functional from Eq. \eqref{functional}.

It is worth to emphasize that the convergence is guaranteed only in the case the field is continuous in time. Nevertheless, we observed convergence in all discretizations of the investigated processes. 

In cases considered below, we implemented the optimization method in Python.

\section{Optimized Charging: a Qubit-Qubit Model}\label{qq}

We start by considering a system composed of a quantum charger charging a quantum battery made of $l$ cells ($l$ qubits), while a classical laser field $\epsilon(t)$ is shining on the charger. The Hamiltonian of the whole system reads ($\hbar=1$)
\begin{equation}\label{Hamiltonian1}
    H=H_A+H_B+H_{AB}-\mu\epsilon(t)\sigma_A^x\otimes\mathbb{I}_B,
\end{equation}
where $H_A=\dfrac{\omega}{2}(-\sigma_A^z+\mathbb{I}_A)\otimes\mathbb{I}_B$ and $H_B=\mathbb{I}_A\otimes\omega(-\sigma_k^z+\mathbb{I}_k)/2$ are the free Hamiltonians of the charger and the battery, respectively, and $H_{AB}=g(\sigma^+_A\sigma^-_{B}+\sigma^-_A\sigma^+_{B})$ the interaction Hamiltonian in which $g$ is the interaction strength. Above, we $\sigma_{x,y,z}$ are Pauli matrices with eigenstates $\sigma_{z}|0\rangle=|0\rangle$, $\sigma_{z}|1\rangle=-|1\rangle$, and we used the notation $\sigma_{\pm}=\sigma_{x}\pm\sigma_{y}$. As we are considering real quantum systems, dissipation is inevitable to occur due to the interaction with the environment.  The master equation reads
\begin{equation}\label{ME}
    \dot{\hat{\rho}}_{AB}=\mathcal{L}\hat{\rho}_{AB}=-i[H, \hat{\rho}_{AB}]+\mathcal{D}_{T}[\hat{\rho}_{AB}],
\end{equation}
where $\mathcal{D}_{T}[\cdot]$ the Lindblad super-operator which accounts for the dissipation and is represented by \begin{equation}\label{dissipator}
    \mathcal{D}_{T}[\hat{\rho}_{AB}]=\gamma(N_b(T)+1)\mathcal{D}_{\sigma_A^-}[\hat{\rho}_{AB}] +\gamma N_b(T)\mathcal{D}_{\sigma_A^+}[\hat{\rho}_{AB}],
\end{equation}  
where 
\begin{equation}
    N_b(T)=\frac{1}{\exp [\omega/(K_B T)] -1},
\end{equation}
$T$ is the temperature of the bath, $\gamma$ fixes the time scale of the dissipation and $\mathcal{D}_b(\cdot)=b(\cdot)b^{\dagger}-1/2\{b^{\dagger}b,\cdot \}$ for any operator $b$. Ground state is taken as the initial state of the charger and the battery: $\hat{\rho}_{AB}^{0}=|00\rangle \langle 00|_{AB}$. The goal is to design a pulse for the laser field such that the final battery state has maximum ergotropy. To do so, the target state is set to $\hat{\rho}_{AB}^{\text{tgt}}=\mathcal{I}_{A}\otimes|1\rangle \langle 1|_{B}$. 

We will compare our results obtained when optimizing the field appearing in Eq. \eqref{Hamiltonian1} with the case of using an oscillatory field interacting with the charger, modelled by the following Hamiltonian \cite{farina2019charger}
\begin{equation}\label{OscillatoryHamiltonian}
    H_{osc}=H_A+H_B+H_{AB}+F(e^{-i\omega t}\sigma_A^+ + e^{i\omega t}\sigma_A^-)\otimes\mathbb{I}_B,
\end{equation}
where $H_A$, $H_B$ and $H_{AB}$ are the same as in Eq. \eqref{Hamiltonian1}, 
and rotating-wave approximation to the interaction between the classical field and the battery was applied. In doing so, we assumed $\omega_{a}\approx \omega$, such that $\omega_{a}+\omega\gg\omega_{a}-\omega$. \blk Based on (\ref{Energy-used2}), the cost of using the oscillatory drive can be calculated (see Appendix \ref{En-of-Fi})  as 
\begin{equation}\label{Energy-used-osc}
    \mathcal{W}_{\tau,osc}= 2\tau+ \frac{1}{\omega}\sin (2\omega \tau).
\end{equation}

\subsection{Efficiency and efficacy boost. }
\begin{figure}
\caption{Optimized 
 single qubit battery charging (in green), compared with a non-optimized charging (in black). In the optimized case, the target state is selected to be the excited state for the battery. The final time is chosen such that the energy and ergotropy extraction achieve their maximum value for the oscillatory field. Parameters used: $N_b(T)=0$, $g=0.2\omega$, $\gamma=0.05\omega$, $F=\mu=0.5\omega$.  }
	\label{qubit-qubit-1cell2}
	\centering
	\begin{subfigure}{0.9\linewidth}
		\includegraphics[width=\linewidth]{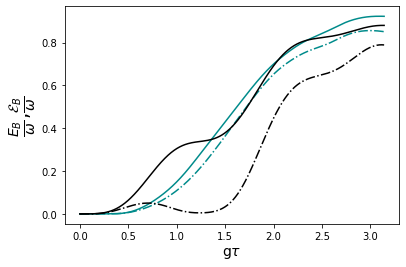}
		\caption{Energy (solid line) and ergotropy (dotted-dashed line) versus time. }\label{fig:subfigA}
	\end{subfigure}
   \vfill
	\begin{subfigure}{0.9\linewidth}
		\includegraphics[width=1\linewidth]{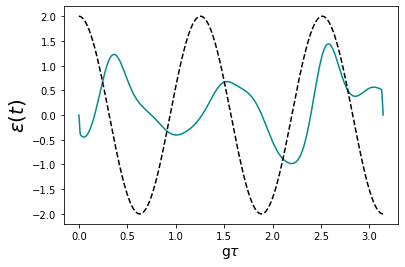}
		\caption{The corresponding field pulses. The initial guess field is taken as $\epsilon^{(i=0)}(t)=S(t)\kappa$, with the amplitude $\kappa=0.5$. }
  \label{fig:subfigB}
	\end{subfigure}

\end{figure}
In Fig. \ref{qubit-qubit-1cell2}, we compare the energy and ergotropy of the battery in both settings. We identify a substantial improvement in terms of energy efficiency, offered by the optimization of the pulse: the energy spent by the optimized field is $\mathcal{W}_{\tau,opt} \approx 6.38$, while the oscillatory drive leads to $\mathcal{W}_{\tau,osc} \approx 31.42$, with quality factors $\alpha_{\mathcal{W}}=392\%$ and $\alpha_{\mathcal{E}}=9.7\%$. We notice that reduced energy expenditure not only is not reflected by the final state of worse quality, but, on the contrary, the state produced is closer to the target in terms of ergotropy, attesting for ability of the optimization method to improve both efficacy and efficiency of the charging process. 
\blk

\subsection{Role of environment in improving efficiency.}\label{Q-Improving}
Moreover, for the qubit-qubit model of the battery charging setup, the optimization method can take advantage of the presence of the above-zero temperature environment to design a pulse which yields improved efficiency over the oscillatory drive. At the same time, not only it consumes significantly less energy than the non-optimized drive, but also less than the optimal pulse for charging in zero-temperature environment.
 
\begin{figure}
\caption{Optimized 
 single qubit battery charging (in green), compared with a non-optimized charging (in black) at time $t=\frac{\pi}{g}$ for different temperatures of the environment. Parameters used $g=0.2\omega$, $\gamma=0.05\omega$, $F=\mu=0.5\omega$, while the non-optimized pulse and initial guess in the optimized scenario are chosen as in Fig 2.}
\label{qubit-temp}
\centering
    \centering
	\begin{subfigure}{0.85\linewidth}
		\includegraphics[width=\linewidth]{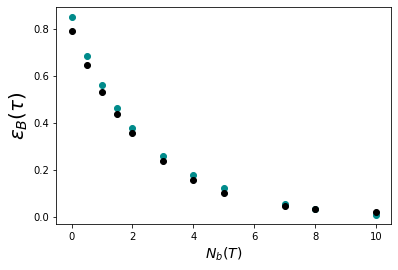}
		\caption{Energy (green) and ergotropy (black) for different temperatures of the environment. Due to the increase of noise, the ergotropy extracted within the optimized approach tends to that provided by the oscillatory field. }\label{fig:subfigA}
	\end{subfigure}
   \vfill
	\begin{subfigure}{0.85\linewidth}
		\includegraphics[width=1\linewidth]{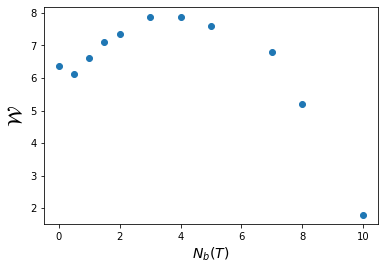}
		\caption{Energy consumed by the optimized field for different temperatures of the environment. Energy spent by the oscillatory field is 31.42 for all temperatures. Energy cost of the optimal pulse grows non-monotonically with small temperatures\blk. 
        }
  \label{fig:subfigB}
	\end{subfigure}
  \vfill

\end{figure}

To show this, in Fig. \ref{qubit-temp} we present ergotropy $\mathcal{E_{B}}$ of the battery resulting from the charging, and energy $\mathcal{W}$ used by the optimized field at time $\tau=\frac{\pi}{g}$ for different temperatures of the environment (energy cost of the oscillatory drive remained 31.42). In order to set comparable energetic constraints for charging in all temperatures, we keep the same choice of $\lambda$ and $\epsilon^{(i=0)}(t)$, given by $(\ref{field-running costs})$, the same as in Fig. \ref{qubit-qubit-1cell2}.

As expected, presence of the constraints results in the inability of the method to compensate for the effect of high temperature noise. Consequently, the value of optimized ergotropy converges to zero, in parallel with the non-optimized behavior.

Nevertheless, temperature dependence of energy spent enables to distinguish three different regimes, as depicted in Fig. \ref{qubit-temp}\textcolor{red}{b}.   First of all, energy consumed by the laser decreases from temperature 0 to some non-zero value ($N_b(T) \approx 1$), which can be explained by feeding energy to the system from the environment, due to the terms of the dissipator which scale with $N_b(T)$. The second trend corresponds to the increasing of energy spent when increasing the average number of photons of the bath $N_b(T)$. This can be explained by the fact that when increasing the temperature, the bath injects more noise into the system, which makes it more difficult for the optimization to get good ratios between ergotropy and energy. Finally, while increasing the temperature further, it becomes impossible for the optimized external field, within the set constraints, to make the system evolve to the target state since, as the system quickly thermalizes to the Gibbs state with zero ergotropy. As a result, the optimization does not ask for energy from the optimized field.
A detailed analysis of the method application to the problem of charging qubit quantum batteries, in particular the role of chosing the optimization parameters in obtaining fast convergence and efficient pulses, is presented in Appendix \ref{AppD}. On the other hand, Appendix \ref{AppE} presents limitations the method faces when applied to charging settings with multiple batteries. It is observed that, within the available computational resources, the method stops offering efficiency and efficacy advantages in charging for settings of 4 and more batteries, connected to a single charger. A further work should be conducted in employing optimization protocols for these settings, for the case they also become accessible in the experiment.\blk

%

\section{Optimized charging: Oscillator-Oscillator Model}\label{oo}
Growing dimensionality of the system subject to optimization can be efficiently tackled in case of linear couplings between harmonic oscillator batteries. In this section, we extend the optimization method to this scenario, taking both the charger and the battery as harmonic oscillators, and subjecting the charger to dissipation. 

Our extension avoids truncating Hilbert space of the systems. The truncation, if employed for optimization of the charging scenario, would produce applicability time limits which become more stringent with increasing coupling to the environment. In contrast, our approach is not susceptible to values of noise, and in principle can always be applied for arbitrary times of the evolution. It is also worth mentioning that from a practical point of view  truncating dimension can become inefficient as one would end up with a large Hilbert space for optimization ( $\mathcal{H}_{oscillator1} \otimes \mathcal{H}_{oscillator2} $) which will suffer from the curse of dimensionality, specially if one wants to store more energy on the battery as it would require higher levels to be populated. Additionally, not having to decide on a dimension size for truncation makes simulation more straightforward and unambiguous. 

The Hamiltonian of our model reads
\begin{equation}\label{Hamiltonian3}
    H=H_A+H_B+H_{AB}-\mu \epsilon(t)(a+a^\dagger)\otimes\mathbb{I}_B,
\end{equation}
where $H_A=\omega a^\dagger a\otimes\mathbb{I}_B$ and $H_B=\mathbb{I}_A\otimes\omega b^\dagger b$ are the free Hamiltonians of the charger and the battery, respectively, and the interaction Hamiltonian reads $H_{AB}=g(ab^{\dagger}+a^{\dagger}b)$. In addition, the charger interacts weakly with the bath. Consequently, the dynamics is described by the Lindblad master Eq. \eqref{ME}, with the Lindblad super-operator accounting for dissipation expressed by
\begin{equation}\label{dissipator}
    \mathcal{D}_T[\hat{\rho}]=\gamma(N_b(T)+1)\mathcal{D}_a[\hat{\rho}]+\gamma N_b(T)\mathcal{D}_{a^\dagger}[\hat{\rho}],
\end{equation}
where $\mathcal{D}_c(\cdot)=c(\cdot)c^{\dagger}-1/2\{c^{\dagger}c,\cdot \}$ for any operator $c$.
The total Hamiltonian $H$ is quadratic, and consequently the evolution preserves the Gaussian character of the initial state. For the initial Gaussian state taken to be the ground state of harmonic oscillator, we can fully describe the state of the total system at any time $t$ \cite{Serafini} using the first and second momenta of the field modes which means that our problem is reduced to optimization over low dimensional space.
We define the following vector of operators
\begin{equation}
    \hat{\textbf{r}}=(\hat{x}_1,\hat{x}_2,\hat{p}_1,\hat{p}_2)^T,
\end{equation}
where $\hat{x}_j=\sqrt{\frac{1}{2\omega}}(\hat{a}_j+\hat{a}_j^{\dagger})$ and $\hat{p}_j=i\sqrt{\frac{\omega}{2}}(\hat{a}_j^{\dagger}-\hat{a}_j)$ with $j=1,2$. The operators of the vector fulfill the commutation relations $[\hat{\textbf{r}},\hat{\textbf{r}}^T]=iJ$, with $J=\left(\begin{array}{cc}
0_2 & \mathbb{I}_2 \\
-\mathbb{I}_2 & 0_2 \\
\end{array} \right )$, where $\mathbb{I}_2$ and $0_2$ are the identity matrix and the null matrix respectively. For the indices $k\in S=\{1,2,3,4\}$ labelling vector elements $\hat{\textbf{r}}_{k}$, the state of the system is fully described by its statistical moments defined through subsets $S'\subset S$:
\begin{eqnarray}
\langle \prod_{k\in S'}\hat{\textbf{r}}_{k}\rangle = \Tr {\hat{\rho}  \prod_{k\in S'}\hat{\textbf{r}}_{k} }
\end{eqnarray}
where $\Pi_{k}$ stands for multiplication 
over $k$. First moments (for $|S'|=1$) and second moments (for $|S'|=2$) fully describe a Gaussian state.  We construct the vector $|r\rangle$ of the first moments 
\begin{equation}
    |r\rangle=\sum_{i=1} ^ 4 \langle\hat{\textbf{r}}_i\rangle |i\rangle,
\end{equation}
as well as the matrix made of the second moments of $x$ and $p$:
\begin{equation}
    V=\frac{1}{2}\sum_{i,j=1} ^ 4 \langle \{ \hat{\textbf{r}}_i, \hat{\textbf{r}}_j \} \rangle |i\rangle \langle j|.
\end{equation}
The relationship between matrix $V$ and the covariance matrix $V_c$ is given by $V_c=V-|r\rangle \langle r|$.  Therefore, in order to apply the optimization method, we will use an object made of the elements of the vector of moments, the elements of the second moment matrix \footnote{As the second moments matrix $V$ is symmetric, we will just use the elements of the upper triangular part of the matrix.} and an independent parameter, which encodes all the information from the vector living in the infinite dimension Hilbert space. We will arrange it as follows
\begin{equation}\label{krotov-vec}
    \psi= \left(\begin{array}{c}
    c \\
     |r\rangle_1 \\
    |r\rangle_2\\
    |r\rangle_3\\
    |r\rangle_4\\
    V_{11} \\
    V_{12} \\
    V_{13} \\
    V_{14} \\
    V_{22} \\
    V_{23} \\
    V_{24} \\
    V_{33} \\
    V_{34} \\
    V_{44}
    \end{array}\right),
\end{equation}
where $c$ is an arbitrary constant to linearize the set of equations; we set it to 1 for simplicity. 
The evolution of our state $\hat{\rho}$ is given by 
\begin{equation}\label{Schrodinger-eq}
    \frac{d\hat{\rho}}{dt}=\mathcal{L}\hat{\rho}=-i[H,\hat{\rho}]+\mathcal{D}_T(\hat{\rho}).
\end{equation}
However, as we are interested in the evolution of the expectation values of the operators, we will resort to the Heisenberg equation. For any operator $O$, the evolution of its expectation value reads
\begin{equation}\label{Heis-eq}
    \frac{d\langle O \rangle}{dt}=i\langle[H,O]\rangle + \langle {\mathcal{D}_T^{\dagger}}(O) \rangle,
\end{equation}
where $\mathcal{D}^{\dagger}_T(O)$ is the adjoint dissipator and is given by 
\begin{eqnarray}\label{adj-diss}
  \mathcal{D}_T^{\dagger}(O)&=&  \gamma (N_b(T)+1) \left(  a^{\dagger} O a - \frac{1}{2} \{a^{\dagger} a ,O\} \right) \nonumber \\&+& \gamma N_b(T) \left(  a O a^{\dagger} - \frac{1}{2} \{a a^{\dagger} , O\}  \right).
\end{eqnarray}
Computing Eq. \eqref{Heis-eq} for all the elements of Eq. \eqref{krotov-vec} we can write the evolution as
\begin{equation}\label{forward-oscillator}
    \dfrac{\partial \psi}{\partial t}=A_f\psi,
\end{equation}
which will be our Schrodinger-like equation of motion with the condition that $\psi(t=0)=\phi_i$, where $\phi_i$ is the initial state of the system rewritten in terms of first and second moments. Consequently, the backwards evolution of the co-states will be given by the adjoint Liouvillian dynamics 
\begin{equation}
    \frac{d\langle O \rangle}{dt}=i\langle[H,O]\rangle + \langle \mathcal{D}_T(O) \rangle,
\end{equation}
where 
\begin{eqnarray}\label{adj-diss}
  \mathcal{D}_T(O)&=&  \gamma (N_b(T)+1) \left( a O a^{\dagger}  - \frac{1}{2} \{a^{\dagger} a ,O\} \right) \nonumber \\&+& \gamma N_b(T) \left(   a^{\dagger} O a - \frac{1}{2} \{a a^{\dagger} , O\}  \right),
\end{eqnarray}
and this will be translated to our Gaussian framework as 
\begin{equation}\label{backwards-oscillator}
    \dfrac{\partial \chi}{\partial t}=A_b\chi,
\end{equation}
where it should be noted that the time runs backwards. The boundary condition in this case is $\chi(\tau)=\phi_t$, where $\phi_t$ is the target state and $\tau$ the final time of our evolution. The explicit forms of $A_f$ and $A_b$ are deferred to Appendix \ref{AppB}. 

Finally, the update of the field will be given by the derivative of the Liouvillian (multiplied by the imaginary unit) with respect to the field only. So, for the iteration $i$ and for the control $\epsilon$ we have
\begin{eqnarray}\label{field-update-oscillators}\nonumber
\Delta \epsilon^{i} (t) &=& \frac{S(t)}{\lambda} \textnormal{Im}\left [ \Tr {\hat{\sigma}^{i-1}  (t) \frac{i\partial \mathcal{L}}{\partial \epsilon(t)}\Bigr|_{(i)}\hat{\rho}^{i} (t)}\right]\\\nonumber
&=&- \frac{S(t)}{\lambda} \textnormal{Im}\left [ \Tr {\hat{\sigma}^{i-1}(t)[\mu(a_1+a_1^{\dagger}),\hat{\rho}^{i} (t)]}\right],\\
\end{eqnarray}
where $\hat{\rho}$ and $\hat{\sigma}$ represent the density matrix of the forward and backward state respectively. This update can be calculated explicitly for any given pair of states and the result is a function of the elements of $\bar{\textbf{r}}$ and $V$ (we refer the Reader to Appendix \ref{AppC} for the details). This makes Eq. \eqref{field-update-oscillators} well defined for any pair of $\psi$ and $\chi$.

\begin{figure}
\caption{Optimized 
 single oscillator battery charging (in green), compared with a non-optimized charging (in black). Energy (solid line) and ergotropy (dashed/dotted line) versus time. Parameters used $g=0.2\omega$, $\gamma=0.01\omega$, $F=\mu=0.1\omega$, and $\kappa=0.01$. The target state we used for the optimization procedure is the pure state $|\psi\rangle=|0\rangle_C |\alpha\rangle_B$ where $|\alpha\rangle_B$ corresponds to a coherent state of the battery with $\alpha=\sqrt{\frac{3}{5}}(1+i)$.
\label{Fig3}}
\label{qubit-Ergotropy-FieldEnergy}
\centering
    \centering
	\begin{subfigure}{0.85\linewidth}
		\includegraphics[width=\linewidth]{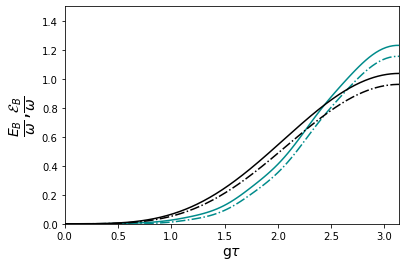}
		\caption{$N_b(T)=1$. Final time chosen such that first maximums in energy and ergotropy for the non-optimized setting appear. }\label{fig:subfigA}
	\end{subfigure}
   \vfill
	\begin{subfigure}{0.85\linewidth}
		\includegraphics[width=1\linewidth]{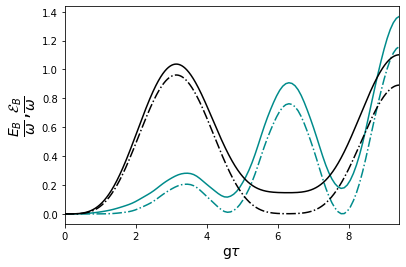}
		\caption{  $N_b(T)=1$. Final time chosen such that second maximums in energy and ergotropy for the non-optimized setting appear.
        }
  \label{fig:subfigB}
	\end{subfigure}
  \vfill
	\begin{subfigure}{0.9\linewidth}
		\includegraphics[width=1\linewidth]{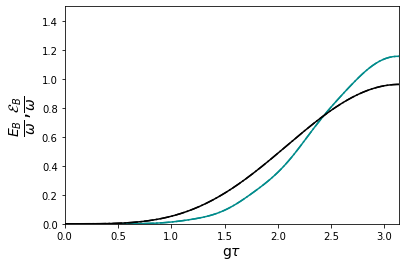}
		\caption{$N_b(T)=0$, and final time as in (a). The state remains pure, with ergotropy and energy taking the values of ergotropy from a).}
  \label{fig:subfigB}
	\end{subfigure}
\end{figure}

Numerical results of our optimization and the non-optimized case are both plotted in Fig. \ref{Fig3}a and it is seen that for the optimized pulse the energy and ergotropy are higher. The energy spent by the optimized laser pulse is 30.73 while by the oscillatory pulse 31.41, resulting in $\alpha_{\mathcal{W}}=2.2\%$.  On the other hand, we observe $\alpha_{\mathcal{E}}= 25\%$, which attests for significant advantage over non-optimized fields in terms of ergotropy extraction. In Fig. \ref{Fig3}b it is seen that for longer times more energy can be transferred to the battery in both optimized and non-optimized process. As longer time offers bigger opportunities for optimization of the field pulse, the energy cost increase is modest (36.7), compared to 94.25 with the fixed sinusoidal pulse ($\alpha_{\mathcal{W}}=157\%$) while there is a sizeable gain in ergotropy extracted $\alpha_{\mathcal{E}}=28\%$. 

One should note that, for initial state of the charger-battery system being $|0\rangle_{C}\otimes|0\rangle_{B}$, the equation of motion (\ref{forward-oscillator}) implies that, for a fixed driving $\epsilon(t)$ at temperature $T$, the ergotropy of the final state does not depend on temperature, following the separation of energy of the target state into parts which depend separately on driving and temperature:
\begin{eqnarray}\label{separation}
\mathcal{E}_{\epsilon(t),T}=E_{\epsilon(t),T=0}=E_{\epsilon(t),T}-E_{\epsilon(t)=0,T}.     
\end{eqnarray}
The proof of Eq. (\ref{separation}) runs analogously to the one in \cite{farina2019charger} for a fixed, sinusoidal driving. It relies on the fact that for quadratic Hamiltonians and a loss channel at temperature $T=0$, the initial vacuum state evolves into a product of coherent states.

From there, it is easy to see how our algorithm could be used in finding optimal driving of a harmonic oscillator battery. Namely, we run the algorithm for target states which have the desired ergotropy, and select the pulse which has the lowest energy cost. Eq. (\ref{separation}) guarantees that this pulse will lead to a state with the same ergotropy when applied at an arbitrary temperature $T$. Simultaneously, a pulse found in this way would also maximize the ratio $\frac{\mathcal{E}}{E}$ for every temperature. 

This properties stand in a striking contrast to the qubit-qubit charging schemes (Section \ref{Q-Improving}), where increasing temperature was connected with diminishing values of extractable ergotropy. As such, it singles out the harmonic oscillator battery architecture as a preferred one when it comes to high tolerance to external noise.

As a closing remark, we note that the method introduced here can be specially useful in fully realistic settings, such as when the initial state cannot be taken as vacuum, or for quadratic interaction with the field.

\section{Realization}
We conclude with the description of the perspective of experimental realisation of the qubit-qubit and oscillator-oscillator models with the current superconducting quantum technology \cite{RevModPhys.93.025005}. 

The architecture is easily scalable to multiple qubit and/or resonator configurations. The chip integration allows coupled multiple transmon qubits or resonator-resonator designs with ultimately stable operations \cite{Arute2019}. The in-situ tunable couplings can be realized using modular capacitive \cite{PhysRevX.4.031043} or inductive \cite{Wulschner2016, PhysRevApplied.16.064041, Wang9} coupling with a large on to off coupling ratio reaching up to GHz ranges. Each qubit or resonator in the chain can also be individually controlled by coupling them to a 1D transmission line or waveguide. Note that, the coupling can be short (capactive) or long (inductive) range, allowing multiple quantum units (resonators or qubits) to efficiently interact; necessary condition to implement the current quantum control proposal. 

Additionally, the microwave control electronics and control methods of these quantum circuits haven been extensively improved and developed over the last few years \cite{PhysRevA.82.040305, PhysRevApplied.12.014044, PhysRevLett.112.240503}. Proper pulse shaping allows faster operations, reducing the gate errors, or improving the gate fidelity. For instance, open or close-loop optimal control methods \cite{Werninghaus2021} allow stable gate operations by optimizing several parameters related to the shape, amplitude, phase, and other properties of the control field acting on the quantum units. 

\section{Conclusions and discussion}
In this paper, we investigated optimization of quantum battery charging, motivated by the fact that, as far as we know, the current literature focuses on simple sinusoidal charging field. Apart from maximising the fidelity with respect to a desired target state and minimising the energy spent, we also took into the account the turning on and off of the field, to make the resulting pulses experimentally feasible. 

Firstly, we studied the case in which both the charger and the battery are two-level systems. In this regime, the optimization procedure returns charging laser pulses which more effectively extract ergotropy on the battery, with a 4-fold increased energetic efficiency. Moreover, the optimization procedure takes properly into account the presence of environment, offering a prospect of further cost reductions of the charging pulses. We also proposed an operational protocol for optimizing the charging process, which should enable determination of an optimal pulse within limited computational resources.

Significant improvements in terms of extracted ergotropy and energy cost were also observed for the case in which harmonic oscillators model both the charger and the battery. Crucially, if product initial conditions are met, in these settings ergotropy extraction does not face challenges linked to finite-temperature effects in charging of qubit-qubit systems, and optimal pulses found for zero temperature are universal. In order to apply the optimization method, it was recast within the Gaussian formalism, making it effective for  systems of infinite dimension.

We advocate that further work in optimizing methods is needed, anticipating the development of experimental techniques enabling fabrication and handling of batteries composed of multiple qubits or harmonic oscillator cells. Namely, for a multi-battery composed of 4 qubit cells, while the application of the current optimization procedure still offers an improved final state energy and ergotropy, the optimized pulse requires more energy, compared with the sinusoidal drive. This is because performing the optimization and at the same time minimizing the energy used by the field is not numerically stable. We conjecture that with higher computational power the optimization could be carried more efficiently, resulting in pulses with lower energy costs.  Moreover, optimization of systems with multiple batteries may benefit from using  gradient-based methods, like  Gradient Ascent
Pulse Engineering (GRAPE) algorithm \cite{khaneja2005optimal}. Note however that these methods are not guaranteed to converge monotonically. Moreover, as suggested in \cite{Goerz_2019}, implementation of the Krotov algorithm in languages other than Python may lead to performance boost of 2-3 orders of magnitude, which could translate into more efficient optimization of the functional.

Furthermore, multi-cell battery systems bring new questions about the role which  geometry of inter-cell couplings may play in effective charging. Therefore, effective optimization method for these scenarios would be of immediate application. The introduced method efficiently  explores possible optimizations for the current state-of-the-art technology, which allows handling of few qubits and harmonic oscillators, working as quantum batteries.

\begin{acknowledgments}
RRR and GS acknowledge Michael Goerz and Daniel Reich for fruitful discussions. BA would like to akcnowledge Tomasz Linowski for his useful arguments. We acknowledge
support from the Foundation for Polish Science through IRAP project co-financed by EU within the
Smart Growth Operational Programme (contract no.2018/MAB/5). S.B. acknowledges funding by the Natural Sciences and Engineering Research Council of Canada (NSERC) through its Discovery Grant, funding and advisory support provided by Alberta Innovates through the Accelerating Innovations into CarE (AICE) -- Concepts Program, and support from Alberta Innovates and NSERC through Advance Grant.
\end{acknowledgments}

\appendix

\section{Energy used by the field}\label{En-of-Fi}
In order to obtain the interaction Hamiltonian $H_I(t)$ in Eq. \eqref{OscillatoryHamiltonian} the classical oscillatory field is assumed to be of the form
\begin{equation}
    \epsilon(t)=E_0\cos{\omega t}=\frac{E_0}{2}(e^{i\omega t}+e^{-i\omega t}),
\end{equation}
where $E_0$ is the constant amplitude of the field. In the dipole approximation the interaction Hamiltonian $H_I(t)$ is then written as \cite{gerry2005introductory}
\begin{align*}\
    H_I(t)&=\mu \epsilon(t)\sigma^x\\\nonumber
    &=\frac{\mu E_0}{2}(e^{i\omega t}+e^{-i\omega t})\sigma^x,\numberthis
\end{align*}
where $\mu$ is the coupling constant and $\sigma^x$ the Pauli operator. Using the equality $\sigma^{x}=\sigma^++\sigma^-$ where $\sigma^{\pm}$ are the raising and lowering operators and applying the RWA approximation we get
\begin{align*}
    H_I(t)&= \frac{\mu E_0}{2}(e^{i\omega t}\sigma^++e^{-i\omega t}\sigma^-)\\\nonumber
    &= F(e^{i\omega t}\sigma^++e^{-i\omega t}\sigma^-),\numberthis
\end{align*}
which is the interaction Hamiltonian in Eq. \eqref{OscillatoryHamiltonian}. Therefore choosing $E_0=2$ we have $F=\mu$. And the energy spent by the oscillatory field is then computed as
\begin{align*}\label{Energy-used}
    \mathcal{W}_{\tau}&= \int_{0}^{\tau} dt\ |\epsilon(t)|^2\\\nonumber
    &= \int_{0}^{\tau} dt\ 4\cos^2(\omega t)\\\nonumber
    &= 2\tau+ \frac{\sin (2\omega \tau)}{\omega}.\numberthis
\end{align*}

\section{Equations of motion}\label{AppB}

The Krotov method for dissipative systems has been studied before, for instance in Refs. \cite{reich_monotonically_2012,koch_controlling_2016}, it has been established that in this case, the forward and backward equations  are (in which we redefined $\hat{\rho}$ as $\rho_F$ and $\hat{\sigma}$ as $\rho_B$ and now we omit hats on the density matrices):
\begin{eqnarray}
    \frac{d\rho_{F}}{dt} = -i \left[ H,\rho_{F} \right] +  \mathcal{D}_T(\rho_{F}) \\
    \frac{d\rho_{B}}{dt} = -i\left[ H,\rho_{B} \right] - \mathcal{D}_T^{\dagger}(\rho_{B})
\end{eqnarray}
Where
\begin{eqnarray}
  \mathcal{D}_T^{\dagger}(\rho)= \sum_{k}  (A_{k}^{\dagger} \rho A_{k} - \frac{1}{2} A_{k}^{\dagger} A_{k} \rho - \frac{1}{2} \rho A_{k}^{\dagger} A_{k} )
\end{eqnarray}
and 
\begin{eqnarray}
  \mathcal{D}_T(\rho)= \sum_{k}  (A_{k} \rho A_{k}^{\dagger} - \frac{1}{2} A_{k}^{\dagger} A_{k} \rho - \frac{1}{2} \rho A_{k}^{\dagger} A_{k} )
\end{eqnarray}
In order to compute the evolution of an observable we compute the so called ``adjoint'' master equation for both the forward and backwards evaluations, which just means computing $\langle O_{i} \rangle= Tr[ O \rho_{i} ]$, where $i=$forward,backwards which results in:
 
 \begin{eqnarray}
    \frac{d\langle O_{F} \rangle }{dt} = i \langle \left[ H,O_{F} \right] \rangle +  \langle \mathcal{D}_T^{\dagger}(O_{F}) \rangle \\
    \frac{d\langle O_{B} \rangle }{dt} = i \langle \left[ H,O_{B} \right] \rangle -  \langle \mathcal{D}_T(O_{B}) \rangle
\end{eqnarray}

Since for Gaussian systems all of the evolution can be quantified using the vector of first moments and the matrix of second moments, we just need to compute the evolution of the observables that appear as the components of the vector \eqref{krotov-vec} which evolves according to the dynamical Eq. \eqref{backwards-oscillator} where $A_f$ is given by:

\begin{widetext}

\begin{eqnarray}
A_f= \tiny
 \left(\begin{array}{ccccccccccccccc}0 & 0 & 0 & 0 & 0 & 0 & 0 & 0 & 0 & 0 & 0 & 0 & 0 & 0 & 0\\0 & - \frac{\gamma}{2} & 0 & 1 & \frac{g}{w} & 0 & 0 & 0 & 0 & 0 & 0 & 0 & 0 & 0 & 0\\0 & 0 & 0 & \frac{g}{w} & 1 & 0 & 0 & 0 & 0 & 0 & 0 & 0 & 0 & 0 & 0\\ \sqrt{2 w} \mu \epsilon(t) & - w^{2} & - g w & - \frac{\gamma}{2} & 0 & 0 & 0 & 0 & 0 & 0 & 0 & 0 & 0 & 0 & 0\\0 & - g w & - w^{2} & 0 & 0 & 0 & 0 & 0 & 0 & 0 & 0 & 0 & 0 & 0 & 0\\\frac{\gamma \left(n + \frac{1}{2}\right)}{w} & 0 & 0 & 0 & 0 & - \gamma & 0 & 2 & \frac{2 g}{w} & 0 & 0 & 0 & 0 & 0 & 0\\0 & 0 & 0 & 0 & 0 & 0 & - \frac{\gamma}{2} & \frac{g}{w} & 1 & 0 & 1 & \frac{g}{w} & 0 & 0 & 0\\0 &  \sqrt{2 w} \mu \epsilon(t) & 0 & 0 & 0 & - w^{2} & - g w & - \gamma & 0 & 0 & 0 & 0 & 1 & \frac{g}{w} & 0\\0 & 0 & 0 & 0 & 0 & - g w & - w^{2} & 0 & - \frac{\gamma}{2} & 0 & 0 & 0 & 0 & 1 & \frac{g}{w}\\0 & 0 & 0 & 0 & 0 & 0 & 0 & 0 & 0 & 0 & \frac{2 g}{w} & 2 & 0 & 0 & 0\\0 & 0 &  \sqrt{2 w} \mu \epsilon(t)  & 0 & 0 & 0 & - w^{2} & 0 & 0 & - g w & - \frac{\gamma}{2} & 0 & \frac{g}{w} & 1 & 0\\0 & 0 & 0 & 0 & 0 & 0 & - g w & 0 & 0 & - w^{2} & 0 & 0 & 0 & \frac{g}{w} & 1\\\gamma w \left(n + \frac{1}{2}\right) & 0 & 0 &  2 \sqrt{2 w} \mu \epsilon(t) & 0 & 0 & 0 & - 2 w^{2} & 0 & 0 & - 2 g w & 0 & - \gamma & 0 & 0\\0 & 0 & 0 & 0 &  \sqrt{2 w} \mu \epsilon(t)  & 0 & 0 & - g w & - w^{2} & 0 & - w^{2} & - g w & 0 & - \frac{\gamma}{2} & 0\\0 & 0 & 0 & 0 & 0 & 0 & 0 & 0 & - 2 g w & 0 & 0 & - 2 w^{2} & 0 & 0 & 0\end{array}\right)
\end{eqnarray}
\end{widetext}

In the same way the matrix that defines the backward evolution is given by:

\begin{widetext}
\begin{eqnarray}
A_{b}= \tiny \left(\begin{array}{ccccccccccccccc}0 & 0 & 0 & 0 & 0 & 0 & 0 & 0 & 0 & 0 & 0 & 0 & 0 & 0 & 0\\0 & - \frac{3 \gamma}{2} & 0 & 1 & \frac{g}{w} & 0 & 0 & 0 & 0 & 0 & 0 & 0 & 0 & 0 & 0\\0 & 0 & - \gamma & \frac{g}{w} & 1 & 0 & 0 & 0 & 0 & 0 & 0 & 0 & 0 & 0 & 0\\- \sqrt{2 w} \epsilon(t) \mu  & - w^{2} & - g w & - \frac{3 \gamma}{2} & 0 & 0 & 0 & 0 & 0 & 0 & 0 & 0 & 0 & 0 & 0\\0 & - g w & - w^{2} & 0 & - \gamma & 0 & 0 & 0 & 0 & 0 & 0 & 0 & 0 & 0 & 0\\- \frac{\gamma \left(n + \frac{1}{2}\right)}{w} & 0 & 0 & 0 & 0 & - 2 \gamma & 0 & 2 & \frac{2 g}{w} & 0 & 0 & 0 & 0 & 0 & 0\\0 & 0 & 0 & 0 & 0 & 0 & - \frac{3 \gamma}{2} & \frac{g}{w} & 1 & 0 & 1 & \frac{g}{w} & 0 & 0 & 0\\0 & - \sqrt{2 w} \epsilon(t) \mu  & 0 & 0 & 0 & - w^{2} & - g w & - 2 \gamma & 0 & 0 & 0 & 0 & 1 & \frac{g}{w} & 0\\0 & 0 & 0 & 0 & 0 & - g w & - w^{2} & 0 & - \frac{3 \gamma}{2} & 0 & 0 & 0 & 0 & 1 & \frac{g}{w}\\0 & 0 & 0 & 0 & 0 & 0 & 0 & 0 & 0 & - \gamma & \frac{2 g}{w} & 2 & 0 & 0 & 0\\0 & 0 & - \sqrt{2 w} \epsilon(t) \mu  & 0 & 0 & 0 & - w^{2} & 0 & 0 & - g w & - \frac{3 \gamma}{2} & 0 & \frac{g}{w} & 1 & 0\\0 & 0 & 0 & 0 & 0 & 0 & - g w & 0 & 0 & - w^{2} & 0 & - \gamma & 0 & \frac{g}{w} & 1\\- \gamma w \left(n + \frac{1}{2}\right) & 0 & 0 & - 2 \sqrt{2 w} \epsilon(t) \mu  & 0 & 0 & 0 & - 2 w^{2} & 0 & 0 & - 2 g w & 0 & - 2 \gamma & 0 & 0\\0 & 0 & 0 & 0 & - \sqrt{2 w} \epsilon(t) \mu  & 0 & 0 & - g w & - w^{2} & 0 & - w^{2} & - g w & 0 & - \frac{3 \gamma}{2} & 0\\0 & 0 & 0 & 0 & 0 & 0 & 0 & 0 & - 2 g w & 0 & 0 & - 2 w^{2} & 0 & 0 & - \gamma\end{array}\right)
\end{eqnarray}
\end{widetext}

One may notice that the difference between both matrices is merely a shift on the diagonal and part of the ``constant'' terms, to see this better consider $M:=A_f-A_b$:

\begin{widetext}
\begin{eqnarray}
\displaystyle M=\left(\begin{array}{ccccccccccccccc}0 & 0 & 0 & 0 & 0 & 0 & 0 & 0 & 0 & 0 & 0 & 0 & 0 & 0 & 0\\0 & \gamma & 0 & 0 & 0 & 0 & 0 & 0 & 0 & 0 & 0 & 0 & 0 & 0 & 0\\0 & 0 & \gamma & 0 & 0 & 0 & 0 & 0 & 0 & 0 & 0 & 0 & 0 & 0 & 0\\0 & 0 & 0 & \gamma & 0 & 0 & 0 & 0 & 0 & 0 & 0 & 0 & 0 & 0 & 0\\0 & 0 & 0 & 0 & \gamma & 0 & 0 & 0 & 0 & 0 & 0 & 0 & 0 & 0 & 0\\\frac{2 \gamma \left(n + \frac{1}{2}\right)}{w} & 0 & 0 & 0 & 0 & \gamma & 0 & 0 & 0 & 0 & 0 & 0 & 0 & 0 & 0\\0 & 0 & 0 & 0 & 0 & 0 & \gamma & 0 & 0 & 0 & 0 & 0 & 0 & 0 & 0\\0 & 0 & 0 & 0 & 0 & 0 & 0 & \gamma & 0 & 0 & 0 & 0 & 0 & 0 & 0\\0 & 0 & 0 & 0 & 0 & 0 & 0 & 0 & \gamma & 0 & 0 & 0 & 0 & 0 & 0\\0 & 0 & 0 & 0 & 0 & 0 & 0 & 0 & 0 & \gamma & 0 & 0 & 0 & 0 & 0\\0 & 0 & 0 & 0 & 0 & 0 & 0 & 0 & 0 & 0 & \gamma & 0 & 0 & 0 & 0\\0 & 0 & 0 & 0 & 0 & 0 & 0 & 0 & 0 & 0 & 0 & \gamma & 0 & 0 & 0\\2 \gamma w \left(n + \frac{1}{2}\right) & 0 & 0 & 0 & 0 & 0 & 0 & 0 & 0 & 0 & 0 & 0 & \gamma & 0 & 0\\0 & 0 & 0 & 0 & 0 & 0 & 0 & 0 & 0 & 0 & 0 & 0 & 0 & \gamma & 0\\0 & 0 & 0 & 0 & 0 & 0 & 0 & 0 & 0 & 0 & 0 & 0 & 0 & 0 & \gamma\end{array}\right).
\end{eqnarray}
\end{widetext}

To see the general form of $M$ (without going element by element), we just need to compute the difference between the dissipative parts of the backward and forward equations.

The dissipator for the backward equation is
\begin{eqnarray}
  \mathcal{D}_T^{\dagger}(O)&=&  \gamma (N_b+1) \left(  a^{\dagger} O a - \frac{1}{2} \{a^{\dagger} a ,O\} \right) \nonumber \\&+& \gamma N_b \left(  a O a^{\dagger} - \frac{1}{2} \{a a^{\dagger} , O\}  \right)
\end{eqnarray}
and for the forward

\begin{eqnarray}
  \mathcal{D}_T(O)&=& \gamma (N_b+1) \left(  a O a^{\dagger} - \frac{1}{2} \{a^{\dagger} a, O\}  \right) \nonumber \\&+& \gamma N_b \left(  a^{\dagger} O a - \frac{1}{2} \{a a^{\dagger}   ,O \} \right).
\end{eqnarray}

Since there is a minus sign in the backwards evolution the difference in the dynamics is given by the sum of these generators:

\begin{align*}\label{L+Ldagger}
  \mathcal{D}_T(O)+\mathcal{D}_T^{\dagger}(O)  &= \gamma N_b \left( a^{\dagger} O a + a O a^{\dagger} - \{a a^{\dagger},O \}\right) \\\nonumber
  &+ \gamma (N_b + 1) \left( a^{\dagger} O a + a O a^{\dagger} - \{ a^{\dagger} a,O \}\right)\numberthis.
\end{align*}

We may rewrite Eq. \eqref{L+Ldagger} by using the equality:
\begin{align*}
    a^{\dagger} O a + a O a^{\dagger} - a a^{\dagger} O - O a a^{\dagger}&= a [ O, a^{\dagger} ] + [a^{\dagger},O a] \\\nonumber
    &= a [ O, a^{\dagger} ] -  [ O,a^{\dagger} ] a + O \\\nonumber
    &= [a, [O, a^{\dagger}]] - O\numberthis.
\end{align*}
Similarly one may obtain:
\begin{align*}
    a^{\dagger} O a + a O a^{\dagger}- a^{\dagger} a O - O a^{\dagger} a\\
    = [a^{\dagger},[O,a]] + O.\numberthis
\end{align*}
One may write:
\begin{eqnarray}
  \mathcal{D}_T(O)+\mathcal{D}_T^{\dagger}(O)  &=& \gamma N_b \left( [a,[O,a^{\dagger}]]-O \right) \nonumber \\ &+& \gamma (N_b+1) \left( [a^{\dagger},[O,a]]+O \right) \nonumber \\ &=&\gamma N_b \left([a,[O,a^{\dagger}]] + [a^{\dagger},[O,a]] \right) \nonumber \\ &+& \gamma \left([a^{\dagger},[O,a]] + O \right)
\end{eqnarray}
This expression helps to construct the backward equation from the forward one. In the $X,P$ basis, this can be written as:
\begin{align*} \label{eq:general_difference}
  \mathcal{D}_T(O)+\mathcal{D}_T^{\dagger}(O)  &= \gamma N_b w \left([X_{1},[O,X_{1}]] +\frac{1}{w^{2}} [P_{1},[O,P_{1}]] \right) \\\nonumber
  &+ \gamma \frac{w}{2} \left([X_{1},[O,X_{1}]] +\frac{1}{w^{2}} [P_{1},[O,P_{1}]] \right. \\\nonumber
  &+ \frac{i}{w}\left( [X_{1},[O,P_{1}]-[P_{1},[O,X_{1}]]\right)\Big)+ \gamma O, \numberthis
\end{align*}
which allows to compute the backwards evolution from the forward straight-forwardly. For instance let us consider $X_{1}$ and $X_{1}^{2}$
\begin{eqnarray}
    [X_{1},P_{1}]&=&i \\
    \left[X_{1}^{2},P_{1}\right]&=&2 i  X_{1} \\
    \left[P_{1},[X_{1},P_{1}]\right]&=&0  \\
    \left[P_{1},[X_{1}^{2},P_{1}]\right]&=&2
\end{eqnarray}
substituting those relations in Eq. \eqref{eq:general_difference} one obtains:
\begin{equation}
  \langle  \mathcal{D}_T(X_{1})+\mathcal{D}_T^{\dagger}(X_{1}) \rangle= \gamma \langle X_{1} \rangle,
\end{equation}
\begin{equation}
  \langle  \mathcal{D}_T(X_{1}^{2})+\mathcal{D}_T^{\dagger}(X_{1}^{2}) \rangle= \gamma \langle X_{1}^{2}\rangle + \frac{2 \gamma}{w}\left(N_B+ \frac{1}{2}\right). 
\end{equation}

\section{Explicit calculation of the trace of Eq. \eqref{field-update-oscillators}}\label{AppC}
Any two-mode Gaussian state can be written as \cite{Serafini}

\begin{equation}\label{Glauber}
\rho=\frac{1}{(2 \pi)^{2}} \int_{\mathbb{R}^{4}} \mathrm{~d} \mathbf{r} \chi_G({\mathbf{r}}) \hat{D}_{\mathbf{r}},
\end{equation}
with the characteristic function
\begin{equation}
   \chi_G({\mathbf{r}})=\mathrm{e}^{-\frac{1}{2} \mathbf{r}^{\top} \Omega^{\top} V_c \Omega \mathbf{r}+i \mathbf{r}^{\top} \Omega^{\top} \overline{\mathbf{r}}}
\end{equation}

and
\begin{equation}
\hat{D}_{\mathbf{r}}=\mathrm{e}^{i \mathbf{r}^{\top} \Omega \mathbf{r}}
\end{equation}
in which $\mathbf{r}=[x_{1},p_{1},x_{2},p_{2}]$, $V_c$ is the covariance
matrix with entries ${V_c}_{\{i,j\}}=\Tr{\rho \Big(\frac{1}{2}(\textbf{r}_{i}\textbf{r}_{j}+\textbf{r}_{j}\textbf{r}_{i})\Big)}-\bar{\mathbf{r}}_i \bar{\mathbf{r}}_j$, and $[\bar{\mathbf{r}}]_{i}=\Tr{\rho \textbf{r}_{i}}$. Defining $\mathbf{r_{i}}=[x_{i},p_{i}]$ we have
\begin{equation}
    \mathbf{r}=\mathbf{r}_{1}\oplus\mathbf{r}_{2},
\end{equation}
which implies
\begin{equation}
    D_{\mathbf{r}}=D_{\mathbf{r}_{1}}\otimes D_{\mathbf{r}_{2}}.
\end{equation}

In the calculation we will also use identity expansion (on the two mode space and also the fact that all our simulations have been done with $\omega=1$) in the basis of one-mode coherent states:
\begin{equation}
    \mathds{1}=\frac{1}{2\pi}\int \mathrm{~d}\tilde{x}_{1}\mathrm{~d}\tilde{p}_{1} |\frac{\tilde{x}_{1}+i\tilde{p}_{1}}{\sqrt{2}}\rangle\langle \frac{\tilde{x}_{1}+i\tilde{p}_{1}}{\sqrt{2}}|\otimes \mathds{1}_{2},
\end{equation}
where we explicitly used a representation of a coherent single mode state $|\alpha\rangle=|\frac{\tilde{x}_{1}+i\tilde{p}_{1}}{\sqrt{2}}\rangle$.

By denoting by $\chi_{G}(\textbf{r}')$ the characteristic function of the state $\sigma$ (see Eq. \eqref{Glauber}), we obtain 
\begin{widetext}
\begin{eqnarray}\label{Tr}
     \Tr(\sigma (a_1+a_1^{\dagger})\rho)=\Tr(\sigma (a_1+a_1^{\dagger})\rho^{\dagger})& &\nonumber\\
    =\dfrac{1}{(2\pi)^{4}} \Tr(\int_{\mathbb{R}^{4}} \mathrm{~d} \mathbf{r}\int_{\mathbb{R}^{4}} \mathrm{~d} \mathbf{r'}\chi^{
    \ast}_{G} (\mathbf{r})\chi_{G}(\mathbf{r'})    D_{\mathbf{r'}}(a_1+a_1^{\dagger}) D_{-\mathbf{r}})&&\nonumber\\
    =\dfrac{1}{(2\pi)^{5}} \int_{\mathbb{R}^{4}} \mathrm{~d} \mathbf{r}\int_{\mathbb{R}^{4}} \mathrm{~d} \mathbf{r'}\chi^{
    \ast}_{G} (\mathbf{r})\chi_{G}(\mathbf{r'}) \int_{\mathbb{R}^{2}}\mathrm{~d}\tilde{x}_{1}\mathrm{~d}\tilde{p}_{1}   \Big(\Tr(D_{\mathbf{r'}}a_1|\alpha\rangle\langle \alpha| D_{\mathbf{-r}})+\Tr(D_{\mathbf{r'}}|\alpha\rangle\langle \alpha|a_1^{\dagger}D_{\mathbf{-r}})\Big)&&\nonumber\\
    =\dfrac{1}{(2\pi)^{5}} \int_{\mathbb{R}^{4}} \mathrm{~d} \mathbf{r}\int_{\mathbb{R}^{4}} \mathrm{~d} \mathbf{r'}\chi^{
    \ast}_{G} (\mathbf{r})\chi_{G}(\mathbf{r'}) \int_{\mathbb{R}^{2}}\mathrm{~d}\tilde{x}_{1}\mathrm{~d}\tilde{p}_{1} \sqrt{2}\tilde{x}_{1}\langle \alpha|D_{\mathbf{-r}_{1}}D_{\mathbf{r'}_{1}}|\alpha\rangle\Tr(D_{\mathbf{-r}_{2}}D_{\mathbf{r'}_{2}}),  &&
    \end{eqnarray}

where we exploited $a_{1}|\alpha\rangle=\frac{x_{1}+ip_{1}}{\sqrt{2}}|\alpha\rangle$.
Upon utilising 
\begin{equation}
    \Tr(D_{\mathbf{-r}_{2}}D_{\mathbf{r'}_{2}})=2\pi\delta(\mathbf{-r}_{2}+\mathbf{r'}_{2})=2\pi\delta(-x_{2}+x'_{2})\delta(-p_{2}+p'_{2}),
\end{equation} 
\begin{equation}
    D_{\mathbf{\alpha}}
|0\rangle=|\alpha\rangle,
\end{equation}
\begin{equation}
    D_{\mathbf{\alpha}}D_{\mathbf{\beta}}=e^{\alpha\beta^{\ast}-\alpha^{\ast}\beta}D_{\mathbf{\beta}}D_{\mathbf{\alpha}},
\end{equation}
\begin{equation}
    \langle\alpha|\beta\rangle=e^{-\frac{|\alpha|^{2}+|\beta|^2-2\alpha^{\ast}\beta}{2}},
\end{equation}

we obtain
\begin{eqnarray}
    \Tr(\sigma (a_1+a_1^{\dagger})\rho)=\dfrac{\sqrt{2}}{(2 \pi)^{4}} \int_{\mathbb{R}^{4}} \mathrm{~d} \mathbf{r}\int \mathrm{~d} x'_{1} \mathrm{~d} p'_{1} \int \mathrm{~d} \tilde{x}_{1} \mathrm{~d} \tilde{p}_{1} \chi_{G} (x_{1},p_{1},x_{2},p_{2})\chi^{
    \ast}_{G}({x'}_{1},{p'}_{1},x_{2},p_{2})\tilde{x}_{1} \langle \alpha|D_{\mathbf{-r}_{1}}D_{\mathbf{r'}_{1}}|\alpha\rangle=\nonumber\\
    \dfrac{\sqrt{2}}{(2 \pi)^{4}} \int \mathrm{~d} x_{1}\mathrm{~d}p_{1}\mathrm{~d}x_{2}\mathrm{~d}p_{2} \mathrm{~d} x'_{1} \mathrm{~d} p'_{1} \mathrm{~d}\tilde{x}_{1} \mathrm{~d}\tilde{p}_{1} \chi_{G} (x_{1},p_{1},x_{2},p_{2})\chi^{
    \ast}_{G}({x'}_{1},{p'}_{1},x_{2},p_{2})\nonumber\\ \tilde{x}_{1}e^{-\frac{1}{2}(\frac{x_{1}^{2}+p_{1}^{2}+{x'_{1}}^{2}+{p'_{1}}^{2}}{2})-x_{1}{x'_{1}}-p_{1}{p'_{1}}+i\Big(x_{1}{p'_{1}}-{x'_{1}}p_{1}+\tilde{x}_{1}({p'_{1}}-p_{1})+\tilde{p}_{1}({x'_{1}}-x_{1})\Big)}.
\end{eqnarray}
\end{widetext}

This integral can be solved analytically  in terms of first moments and the elements of the covariance matrix $V_c$ of the two states. 

\section{Single qubit battery charging: finding the optimal pulse}\label{AppD}

Below we investigate the relation between parameter $\lambda$ in Eq. (\ref{field-running costs}) and convergence of the algorithm, and suggest a protocol for effective use of the optimization method.

\begin{figure}[t]
\includegraphics[width=0.5\textwidth]{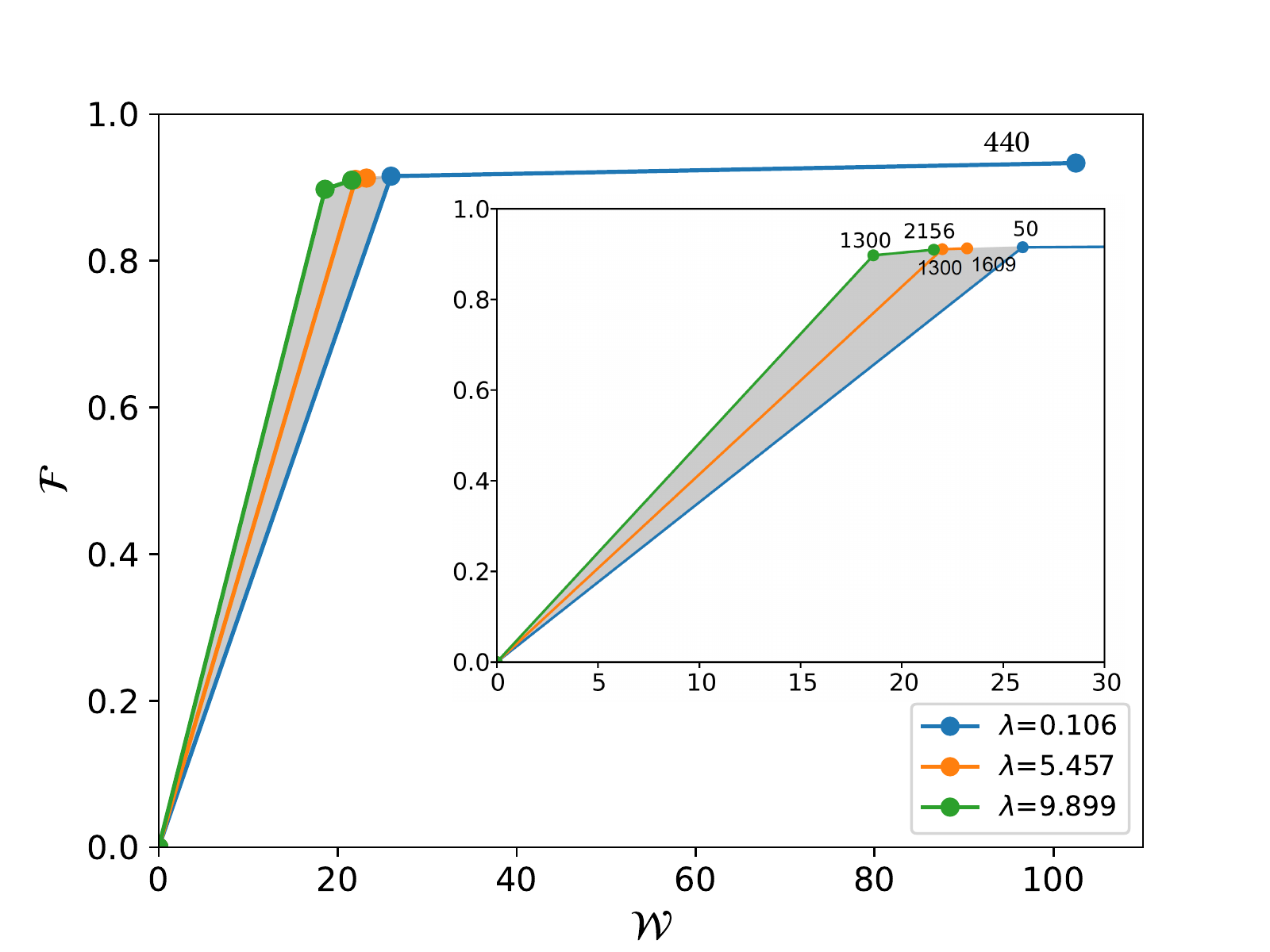}
 \caption{Performance of the algorithm in terms of number of steps, fidelity $\mathcal{F}$ of the final state and energy cost $\mathcal{W}$ of the laser pulse used to  charge a single qubit battery to an excited state. Numbers at points indicate number of steps in which the pulse was obtained. Different lines correspond to different values of $\lambda$ parameter taken.\label{Initial}\label{EW}}
\end{figure}

For the single qubit battery charging, we investigate Pareto fronts on Fidelity vs. Work Cost diagrams, associated with running the algorithm with different $\lambda$ parameters (Fig. \ref{EW}). No strong correlation between fidelity of the required transformation and energy of the pulse is observed. Nevertheless, the algorithm performance depends strongly on $\lambda$. Setting it to higher values forces the algorithm to run for more number of steps before converging, while in the convergence limit it produces a pulse which may be significantly less energetic. From the above observations we infer a recommendation about how the method should be used in practise. Namely, 

1. Running the cost-effective protocol in small $\lambda$ regime to establish the achievable saturation level of fidelity, should be followed by \\

2. Running the cost-demanding protocol in high $\lambda$ regime for number of steps needed to achieve or approach the saturation level from step 1.\\

In this way, the pulse of low energy and leading to high fidelity can be determined even without the need of observing convergence of the method in the cost-demanding regime. The suggested method remains in line with the limit which the optimization problem takes when $\lambda\rightarrow 0$, when energy cost of the pulse is entirely disregarded. Therefore, we expect the proposed approach to be valid for all implementations, though precise values of $\lambda$ must depend on a problem in hand. We also note here that very small values of $\lambda$ which may be chosen in step 1 typically result in strong fluctuations of the pulse in time domain. However, this pulse serves only as a benchmark, and therefore we do not require it to be experimentally feasible. 
\newpage
\blk

\begin{figure}[h!]
    \centering
\caption{Optimized 
 three qubit battery charging (in green), compared with a non-optimized charging (in black). In the optimized case, the target state is selected to be the excited state for the battery. The final time is chosen such that the energy and ergotropy extraction achieve their maximum value for the oscillatory field. Parameters used: $N_b(T)=0$, $g=0.2\omega$, $\gamma=0.05\omega$, $F=\mu=0.5\omega$.  }
	\label{qubit-qubit-1cell}
	\centering
	\begin{subfigure}{0.9\linewidth}
		\includegraphics[width=\linewidth]{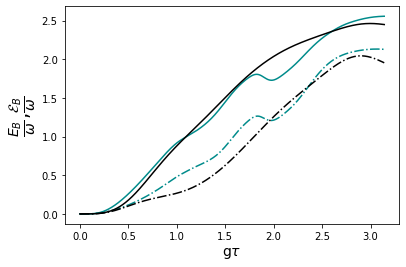}
		\caption{Energy (solid line) and ergotropy (dotted-dashed line) versus time. }\label{fig:subfigA}
	\end{subfigure}
   \vfill
	\begin{subfigure}{0.9\linewidth}
		\includegraphics[width=1\linewidth]{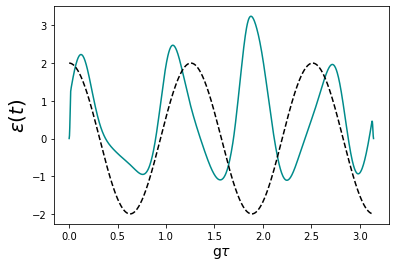}
		\caption{The corresponding field pulses. The initial guess field is taken as $\epsilon^{(i=0)}(t)=S(t)\kappa$, with the amplitude $\kappa=0.75$. }
  \label{fig:subfigB}
	\end{subfigure}

\end{figure}

\section{Multiple qubit battery charging: algorithm performance}\label{AppE}
Below, we analyze the setting with single qubit charger and multiple qubit batteries, expressed as in (\ref{Hamiltonian1}), with the modifications  $H_B=\mathbb{I}_A\otimes\sum_{k=1}^{l}\omega(-\sigma_k^z+\mathbb{I}_k)/2$ and $H_{AB}=\sum_{k=1}^{l}g(\sigma^+_A\sigma^-_{B,k}+\sigma^-_A\sigma^+_{B,k})$, where summations over index $l$ account for the presence of many qubit battery cells which do not interact with each other, but solely with the qubit charger.

For the case of 3 cells, more energy is used by the optimized field compared to the 1 qubit cell, yet it remains below the energy used by the non-optimized field. The quality factors read $\alpha_{\mathcal{W}}=13.8\%$ and  $\alpha_{\mathcal{E}}=8.4\%$, calculated for fixed time $\tau=\frac{\pi}{g}$. The time of achieving maximal ergotropy by a sinusoidal drive decreases when more cells are added. For a 3 qubit cell, it shifts to approximately $\tau=0.92 \frac{\pi}{g}$. With the optimization algorithm ran and compared with the sinusoidal drive for this updated time, it brings a slightly modified improvement factors $\alpha_{\mathcal{W}}=15.93\%$  and $\alpha_{\mathcal{E}}=7.75\%$. Search over different initial gueses of the field profile given by Eq. (\ref{S}) and over the parameter $\lambda$ did not allow us to further improve these factors within the available computational resources. We observed that the method ceases to bring substantial improvements for 4 qubit cell systems, suggesting a need for further improvements in the optimization.

\clearpage

\bibliography{References}

\end{document}